\documentclass[amsmath, amssymb, aps, prb, reprint, floatfix, superscriptaddress]{revtex4-2}

\usepackage[T1]{fontenc}
\usepackage[utf8]{inputenc}
\usepackage{dsfont}
\usepackage{graphicx}
\usepackage{xcolor}
\definecolor{linkscolor}{RGB}{10,55,130}
\usepackage[pdftex, colorlinks=true, linkcolor=linkscolor, citecolor=linkscolor, urlcolor=linkscolor, breaklinks=true]{hyperref}
\usepackage{bm}
\usepackage{braket}
\usepackage{times}
\usepackage{physics}
\usepackage{units}
\usepackage{jabbrv}

\begin{document}

\title{Enhanced localization length in a disordered one-dimensional band\\ via cavity coupling to delocalized states}

\author{F. Mattiotti}
\affiliation{University of Strasbourg and CNRS, CESQ and ISIS (UMR 7006), aQCess, 67000 Strasbourg, France}
\affiliation{Theoretical Physics, Saarland University, D-66123 Saarbrücken, Germany}
\author{G. Pupillo}
\affiliation{University of Strasbourg and CNRS, CESQ and ISIS (UMR 7006), aQCess, 67000 Strasbourg, France}
\affiliation{Institut Universitaire de France (IUF), 75000 Paris, France}
\author{J. Dubail}
\affiliation{University of Strasbourg and CNRS, CESQ and ISIS (UMR 7006), aQCess, 67000 Strasbourg, France}
\author{D. Hagenmüller}
\affiliation{University of Strasbourg and CNRS, IPCMS (UMR 7504), 67000 Strasbourg, France}

\begin{abstract}
We investigate the localization properties of cavity-coupled electronic states in disordered systems, motivated by recent proposals of cavity-mediated hopping in quantum Hall systems. We first introduce a minimal two-band model in which localized states in a disordered one-dimensional band are coupled, through a homogeneous cavity mode, to an excited band of delocalized states. Combining perturbation theory with a transfer-matrix approach, we show that cavity-assisted hopping between localized states decays exponentially with distance, implying that the eigenstates remain localized even beyond the perturbative regime. Nevertheless, the corresponding localization length increases with the light--matter coupling strength and can extend over several lattice sites in the single-electron ultrastrong-coupling regime. We then study a disordered Landau band coupled to a cavity mode within the framework developed in Refs.~\cite{ciuti_cavity-mediated_2021,Appugliese2022barnesreakdown}. We find that the effective cavity-mediated coupling between edge states also decays exponentially with distance, but with a localization length that can reach micrometer scales for experimentally realistic parameters. By analyzing the inverse participation ratio, we show that this enhanced coupling is predominantly mediated by the most extended states of the upper Landau band. Our results demonstrate that, while cavity-induced hopping in disordered quantum Hall systems remains exponentially localized, the associated localization length can become sufficiently large for the corresponding states to exhibit effectively delocalized behavior on mesoscopic length scales.
\end{abstract}

\maketitle

\section{Introduction}

Strong light--matter interactions in photonic structures has emerged as a powerful means of tailoring material properties~\cite{RevModPhys.91.025005,Kockum2019UltrastrongCoupling,garcia-vidal_manipulating_2021,Schlawin2022}, enabling control of energy transfer~\cite{Andrews_2000,coles_polaritonmediated_2014,zhong_energy_2017,Schafer_2019,georgiou_ultralongrange_2021,DelPo_2021,Castagnola_2024} and enhancing exciton~\cite{Feist_2015,Schachenmayer2015,balasubrahmaniyam_enhanced_2023,Fowler-Wright2025Mapping,sandik_cavityenhanced_2025} and charge transport~\cite{orgiu_conductivity_2015,hagenmuller_2017,PhysRevB.97.205303,doi:10.1021/acsnano.0c03496}. Recent work has also explored cavity-induced modifications of topological phases~\cite{PhysRevB.99.235156,DmytrukSchiro2022Controlling,D2CP01806C,PhysRevB.108.245417,NguyenArwasLinYaoCiuti2023ElectronPhotonChern,Bacciconi2024,PhysRevB.109.155160,PerezGonzalez2025lightmatter,Zhao2025cavity}. Because these phases rely on boundary-localized edge states, the nonlocal character of cavity fields raises the possibility that cavity-mediated long-range interactions could affect their robustness. Related long-range effects appear in Kitaev chains, where extended pairing yields gapped phases with algebraic correlations~\cite{PhysRevLett.113.156402}. In the quantum Hall effect, a paradigmatic topological phase, edge states enable robust, backscattering-immune transport arising from a nonzero Chern number~\cite{PhysRevLett.45.494,PhysRevLett.49.405,halperin_quantized_1982,RevModPhys.82.3045}.
 
The coupling of quantum Hall systems to terahertz cavity fields was first proposed about a decade ago~\cite{PhysRevB.81.235303,doi:10.1126/science.1216022}, showing that the ultrastrong coupling regime, where the light--matter coupling becomes a sizable fraction of the transition frequency, can be reached at moderate magnetic fields and high electron densities. In this regime, Shubnikov--de Haas oscillations were predicted~\cite{BartoloCiuti2018VacuumDressed} and later observed to be modified~\cite{Paravicini-Bagliani2019MagnetoTransport}. More recent theory and experiments have demonstrated cavity-induced changes to the transverse Hall response in both integer~\cite{ciuti_cavity-mediated_2021,ArwasCiuti2023Quantum,PhysRevLett.131.196602,Rokaj2022,Appugliese2022barnesreakdown,Enkner2024Testing} and fractional quantum Hall systems~\cite{WinterZilberberg2023FractionalEdgePolaritons,Bacciconi2025_frac,Enkner2025Tunable} under strong magnetic fields.

In the original proposal of cavity-modified Hall conductance~\cite{ciuti_cavity-mediated_2021}, and in subsequent experiments~\cite{Appugliese2022barnesreakdown}, the breakdown of the integer quantum Hall effect was attributed to cavity-mediated hopping between localized states in disordered Landau levels~\cite{halperin_quantized_1982}. In this framework, counter-rotating light--matter coupling terms induce virtual transitions to extended higher-LL states, thereby generating effective long-range hopping within the disordered band, which thus provides a new backscattering channel where two edge states can be connected through the bulk extended states thereby breaking the topological protection. This mechanism has subsequently been explored in various settings~\cite{ArwasCiuti2023Quantum,BoriciArwasCiuti}. While the effective hopping process was analyzed in detail in the original proposal~\cite{ciuti_cavity-mediated_2021}, the nature of its spatial decay has remained elusive. In particular, it is unclear whether the induced hopping decays exponentially with distance, consistent with cavity-hybridized disordered states remaining localized, or instead follows a power law indicative of genuinely long-range processes. The latter scenario would imply algebraically localized states, as encountered for example in power-law random banded matrices~\cite{Mirlin1996,Evers2008}, at the critical point of Anderson metal--insulator transitions~\cite{Wegner1980IPR,Rodriguez2011}, or more recently in random spin models coupled to cavity modes, such as the disordered Tavis--Cummings model~\cite{Dubail2022,Mattiotti2024}.


In this work, we investigate the spatial structure of cavity-mediated long-range hopping and the localization properties of cavity-coupled electronic states. To this end, we introduce a minimal model consisting of localized disordered electronic states in a one-dimensional chain coupled, via a homogeneous cavity mode, to an excited band of delocalized states. This model extends, by including disorder, the one studied in Refs.~\cite{hagenmuller_2017, Hagenmuller_PRB2018}, where the light--matter coupling was shown to enhance conductance by opening an additional transport channel through an excited band. Here, we show that the light--matter interaction generates a hierarchy of hybrid states with increasing photon number, such that in the limit of infinitely many photon sectors the model acquires an effective two-dimensional character. A central question is whether this emergent dimensionality can drive a transition from initially fully localized states to delocalized ones. By combining perturbative arguments with a transfer-matrix analysis, we show that the eigenstates remain exponentially localized across the entire parameter regime considered. However, the corresponding localization length grows with increasing light--matter coupling strength and can extend over several lattice sites in the single-electron ultrastrong-coupling regime, where the coupling exceeds the smallest spectral gap.

Because disorder-broadened LLs comprise spatially localized states in the band tails and extended states near the band centers~\cite{halperin_quantized_1982}, cavity-mediated long-range hopping in our minimal model is naturally connected to the corresponding mechanism in disordered LLs coupled to a cavity mode. Motivated by this analogy, we perform exact numerical simulations of a finite-size disordered Landau band using the model introduced in Ref.~\cite{ciuti_cavity-mediated_2021}. From the resulting eigenstates, we extract the effective hopping amplitudes between states within the disordered Landau band and show that they decay exponentially with distance, as expected for localized systems. Remarkably, however, the associated localization length can substantially exceed the magnetic length and reach micron scales for experimentally realistic parameters. By further analyzing the inverse participation ratio, we demonstrate that the states providing the dominant contribution to cavity-mediated hopping between opposite edges of the quantum Hall bar in the lowest Landau band are precisely the most extended states of the upper Landau band.

Our results can be connected to the theoretical analysis of Ref.~\cite{orgiu_conductivity_2015}, which considered a one-dimensional two-orbital chain with off-diagonal disorder in the intersite hopping in addition to onsite diagonal disorder and found that coupling to a cavity mode can partially delocalize the electronic wave functions. Extending our approach may provide further insight into the mechanism by which coupling to the cavity vacuum field enhances charge transport through the emergence of spatial coherence in regimes that would otherwise exhibit incoherent transport~\cite{orgiu_conductivity_2015,Balasubrahmaniyam2023,KumaretalJACS2024}.

The paper is organized as follows. In Sec.~\ref{oneDmodel}, we introduce a minimal two-band model designed to investigate the localization properties of cavity-coupled electronic states. The analysis combines perturbative arguments (Sec.~\ref{pert}), exact diagonalization (Sec.~\ref{exact_diag}), and a transfer-matrix approach (Sec.~\ref{transfer_mat}) that enables the study of large system sizes. In Sec.~\ref{Landau_band}, we consider a natural realization of the minimal model, namely a disordered Landau band in a quantum Hall system coupled to a single-mode cavity, following Ref.~\cite{ciuti_cavity-mediated_2021}. Using exact diagonalization, we compute the effective hopping amplitudes and the inverse participation ratio, demonstrating that the localization length of edge states can reach mesoscopic length scales. Conclusions and perspectives are presented in Sec.~\ref{conc}.    

\section{Two-band model}
\label{oneDmodel}

\begin{figure}[htbp] 
\centering
\includegraphics[width=0.9\linewidth]{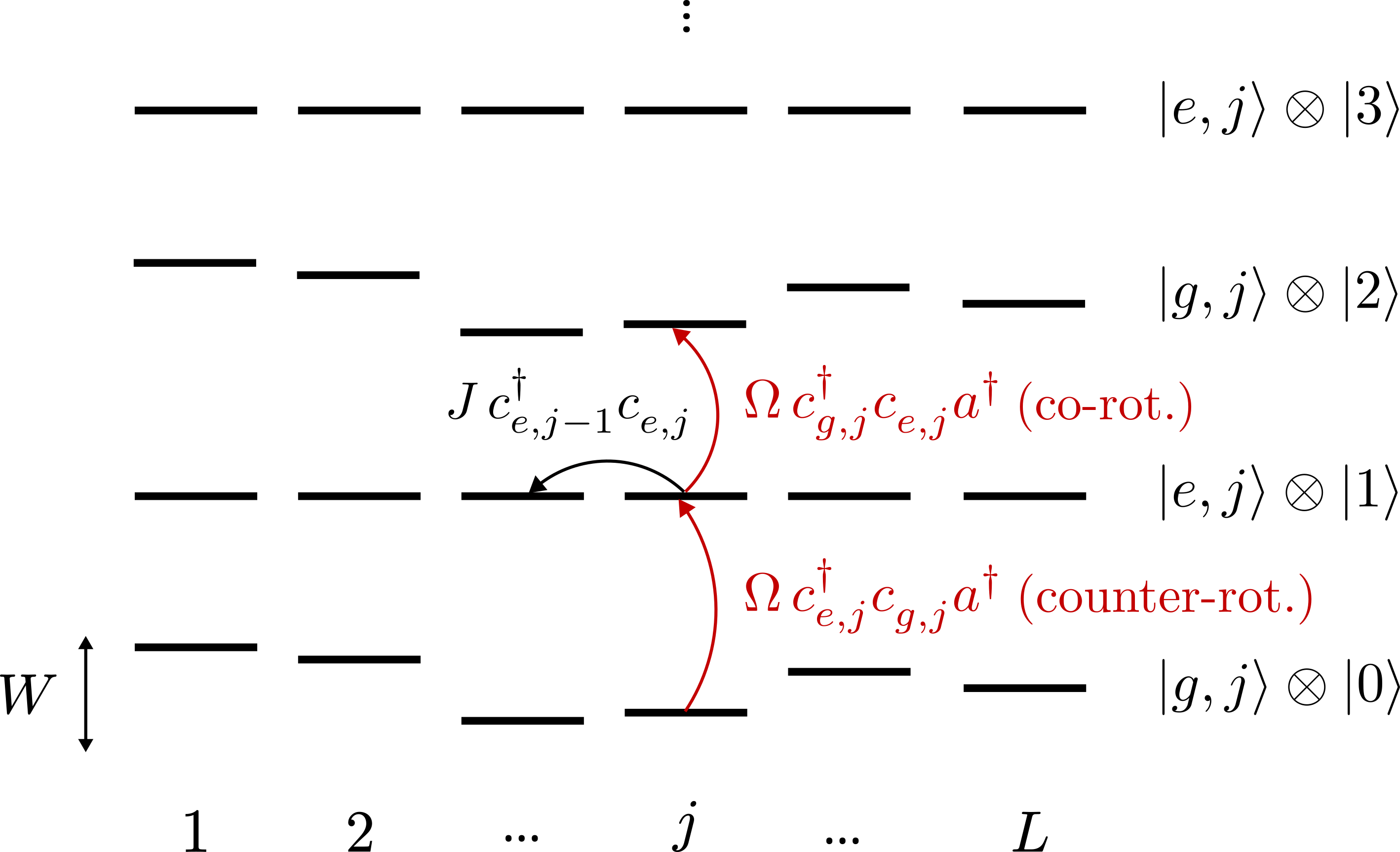}
\caption{\textbf{Schematic of the two-band model.} A disordered lower band of localized states (energies $\epsilon_{g,j}$, width $W$) is coupled via a single-mode cavity to a delocalized upper band with hopping $J$. Light--matter interaction induces virtual transitions between bands: counter-rotating terms couple even- to odd-photon sectors (e.g., $\ket{g,j}\otimes\ket{0} \leftrightarrow \ket{e,j}\otimes\ket{1}$), while co-rotating terms couple odd- to even-photon sectors (e.g., $\ket{e,j}\otimes\ket{1} \leftrightarrow \ket{g,j}\otimes\ket{2}$).}
\label{fig:sketch1}
\end{figure}

We first consider a one-dimensional lattice featuring two electronic orbitals coupled to a single-mode cavity field. The lower-energy band consists of states localized at lattice site $j$, each with an on-site energy $\epsilon_{g,j}$ drawn randomly from a uniform distribution of width $W$, which characterizes the disorder strength. The upper (excited) band comprises electronic states with energy $\epsilon_{e}$ at each site $j$, where electrons hop between nearest neighbors with amplitude $J$ (see Fig.~\ref{fig:sketch1}). The total Hamiltonian (with $\hbar = 1$) is written as $H = H_{0} + V$, where $H_{0} = H_{\mathrm{el}} + H_{\mathrm{ph}}$ describes the uncoupled electronic and photonic degrees of freedom. The photonic Hamiltonian is $H_{\mathrm{ph}} = \omega a^\dagger a$,
where $a^\dagger$ ($a$) denotes the creation (annihilation) operator of a photon with energy $\omega$. The electronic Hamiltonian is given by $H_{\mathrm{el}}=\sum_{j=1}^{L} H^{(j)}_{\mathrm{el}}$, where
\begin{align*}
H^{(j)}_{\mathrm{el}} = \epsilon_{g,j}\, c^\dagger_{g,j} c^{\phantom{\dagger}}_{g,j} 
+ \epsilon_{e}\, c^\dagger_{e,j} c^{\phantom{\dagger}}_{e,j}
+ J \left( c^\dagger_{e,j+1} c^{\phantom{\dagger}}_{e,j} + \textrm{H.c.} \right)~,
\end{align*}
and $c^\dagger_{g,j}$ ($c_{g,j}$) and $c^\dagger_{e,j}$ ($c^{\phantom{\dagger}}_{e,j}$) are the creation (annihilation) operators for electrons in the disordered and excited bands, respectively, localized at site $j$. The coupling between the two bands mediated by the cavity field is described by the interaction Hamiltonian
\begin{align}
V =\Omega \sum_{j=1}^L \left( c^\dagger_{e,j} c^{\phantom{\dagger}}_{g,j} + c^\dagger_{g,j} c^{\phantom{\dagger}}_{e,j} \right) \left( a^\dagger + a \right)~,
\label{eq_coupl}
\end{align}
where $\Omega$ denotes the \textit{on-site} light--matter coupling strength. The coupling Hamiltonian $V$ contains two distinct classes of terms: the co-rotating contributions $\propto c^\dagger_{e,j} c_{g,j} a$ and $\propto c^\dagger_{g,j} c_{e,j} a^{\dagger}$, which conserve the total number of excitations $n_{e}=\sum_{j} c^\dagger_{e,j}c_{e,j}+a^{\dagger}a$ and are retained within the rotating-wave approximation (RWA), and the counter-rotating terms $\propto c^\dagger_{e,j} c_{g,j} a^{\dagger}$ and $\propto c^\dagger_{g,j} c_{e,j} a$, which instead couple sectors with different excitation numbers while preserving the excitation-number parity $(-1)^{n_e}$. This two-band model maps onto an effective 2D lattice, with one axis representing the site index $j$ and the other the cavity photon number (Fig.~\ref{fig:sketch1}). The state $\ket{g,j}\otimes\ket{0}$ (disordered band with zero photons) is coupled, via the counter-rotating terms to the state $\ket{e,j}\otimes\ket{1}$ (excited band with one photon). The latter is coupled both to its nearest neighbors $\ket{e,j\pm1}\otimes\ket{1}$ within the same band and to the state $\ket{g,j}\otimes\ket{2}$ (disordered band with two photons) through the co-rotating terms. This generates a ladder of hybrid light-matter states with increasing photon number.

\subsection{Perturbation theory}
\label{pert}

It is instructive to first evaluate the effective coupling between two localized states in the lower band arising from virtual transitions to the excited band mediated by the cavity field using second-order perturbation theory. Such an approach is valid provided that the light--matter coupling remains small compared to the energy gap between the lower band and the first excited band. To this end, it is convenient to diagonalize the excited band by introducing the Fourier modes $\tilde{c}_{e,k} = \frac{1}{\sqrt{L}} \sum_{j=1}^L e^{2\pi i k j / L} c_{e,j}$, in terms of which the electronic and coupling Hamiltonians take the form
\begin{align}
H^{(j)}_{\mathrm{el}}&=\epsilon_{g,j}\, c^\dagger_{g,j} c^{\phantom{\dagger}}_{g,j} 
+ \sum_{k} \left[\epsilon_{e} + 2J\cos \left( \frac{2\pi k}{L} \right) \right] \tilde{c}^{\dagger}_{e,k} \tilde{c}^{\phantom{\dagger}}_{e,k} \nonumber \\
V&= \frac{\Omega}{\sqrt{L}} (a + a^\dagger) \sum_{j,k} \left( e^{\frac{2\pi \mathrm{i} k j}{L}} \tilde{c}^{\dagger}_{e,k} c^{\phantom{\dagger}}_{g,j} + e^{\frac{-2\pi \mathrm{i} k j}{L}} c^\dagger_{g,j} \tilde{c}^{\phantom{\dagger}}_{e,k} \right)~.
\end{align}
Considering the initial state $\vert j \rangle = c^{\dagger}_{g,j}\ket{\mathrm{VS}}$ and the final state $\vert f \rangle = c^{\dagger}_{g,f}\ket{\mathrm{VS}}$, where $\ket{\mathrm{VS}}$ denotes the vacuum state, the two are coupled through the counter-rotating terms in the interaction Hamiltonian~\eqref{eq_coupl} via intermediate states of the form $a^\dagger c^{\dagger}_{e,j}\ket{\mathrm{VS}}$. In the thermodynamic limit $L \to \infty$, the resulting effective coupling between the initial and final states is given by
\begin{align}
    \label{eq:sum}
    \Gamma_{f,j} &= \bra{f} V \frac{1}{\epsilon_{g,j} - H_0} V \ket{j} \nonumber \\
    &= \frac{\Omega^2}{L} \sum_{k} \frac{e^{2\pi \mathrm{i} k (j-f)/L}}{\epsilon_{g,j} - \epsilon_{e} - 2J\cos \left( \frac{2\pi k}{L} \right) - \omega}~.
\end{align}
In the thermodynamic limit $L\to \infty$, the sum can be replaced by an integral over the continuous variable $q=2\pi k/L$:
\begin{equation}
    \Gamma_{f,j} = \frac{\Omega^2}{2\pi} \int_0^{2\pi} dq \frac{e^{\mathrm{i} q(j-f)}}{\epsilon_{g,j} - \epsilon_{e} - 2J \cos (q) - \omega}~.
\end{equation}
It is convenient to introduce the parameter $\Delta_j = \epsilon_{g,j} - \epsilon_{e} - \omega$, whose magnitude corresponds to the lowest spectral gap, namely the energy separation between states in the disordered band and states in the excited band dressed by one photon, together with the complex variable $z = e^{\mathrm{i} q}$. The integration contour then becomes the unit circle $C(1)$ in the complex plane, yielding
\begin{equation}
    \Gamma_{f,j} = -\frac{\Omega^2}{2\pi \mathrm{i} J} \oint_{C(1)} dz \frac{z^{j-f}}{z^2 - \frac{\Delta_j}{J} z + 1}~.
\end{equation}
The integrand possesses two simple poles,
\begin{equation}
    z_\pm = \frac{\Delta_j}{2J} \pm \sqrt{\frac{\Delta_j^2}{4J^2} - 1}~.
\end{equation}
Moreover, for $j>f$ the integrand has a pole of order $j-f$ at $z=\infty$, whereas for $j<f$ it has a pole of order $f-j$ at $z=0$. In the following, we focus on the \textit{narrow-bandwidth regime} $2J \ll |\Delta_j|$, where the width of the excited band is small compared to the other relevant energy scales.

In this limit, both poles $z_\pm$ are real. Let us first consider the case $j \geq f$. A single pole, namely $z_-$ for $\Delta_j>0$ (or $z_+$ for $\Delta_j<0$), lies inside $C(1)$. Applying the residue theorem then gives
\begin{equation}
    \Gamma_{f,j} = -\frac{\Omega^2}{J} \frac{z_-^{j-f}}{z_- - z_+}.
\end{equation}
For $\Delta_j<0$, one should exchange $z_+ \leftrightarrow z_-$. Next we consider the case $j<f$. The integrand contains an additional pole at $z=0$, together with a single pole outside $C(1)$, located at $z=z_+$ for $\Delta_j>0$ (or $z=z_-$ for $\Delta_j<0$). Applying the residue theorem outside the contour therefore yields
\begin{equation}
    \Gamma_{f,j} = -\frac{\Omega^2}{J} \frac{z_+^{j-f}}{z_- - z_+}~.
\end{equation}
Again, for $\Delta_j<0$ one exchanges $z_+ \leftrightarrow z_-$.

\begin{figure}[htbp!]
    \centering
    \includegraphics[width=0.9\linewidth]{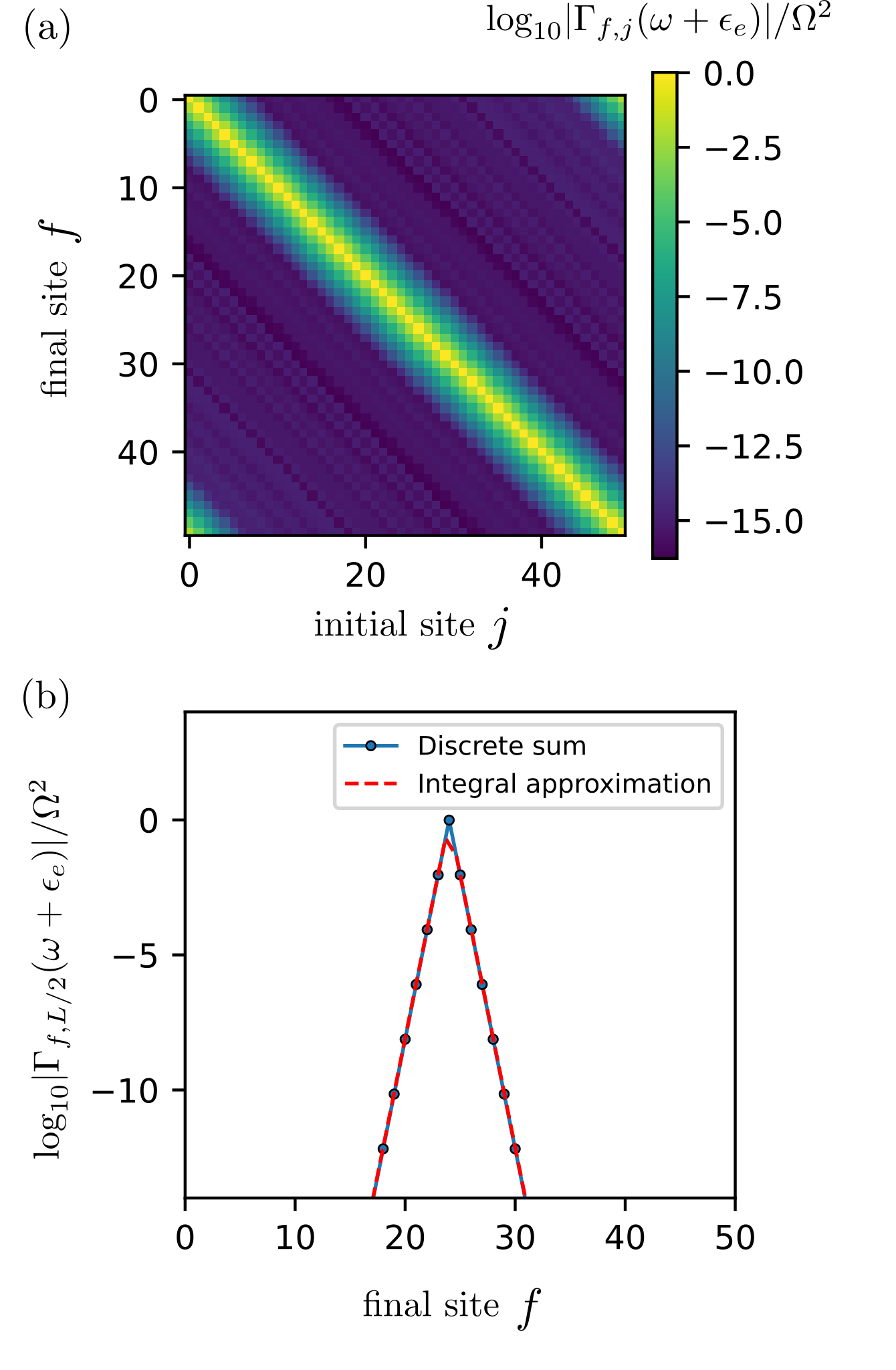}
    \caption{\textbf{Cavity-mediated coupling in the perturbative regime.} Perturbative cavity-mediated coupling between the initial state $\ket{g,j}$ and the final state $\ket{g,f}$, both in the disordered band (logarithmic scale). The coupling is evaluated using Eq.~\eqref{eq:sum} [displayed as the 2D colormap in (a) and blue markers in (b)] and compared with the approximate analytical expression of Eq.~\eqref{eq:integral} [red dashed curve in (b)]. Parameters: $\epsilon_{g,j}\in[-0.5,0.5]$ uniformly random distributed, $J=0.1$, $\Omega=0.01$, $\omega+\epsilon_{e}=3$, and $L=50$.}
    \label{fig:coupling}
\end{figure}

Expanding the poles for $J/|\Delta_j| \ll 1$, we obtain 
\begin{equation}
    z_\pm \approx \frac{\Delta_j}{2J} \pm \frac{|\Delta_j|}{2J} \mp \frac{J}{|\Delta_j|}~,
\end{equation}
which leads to (both for $j\geq f$ and $j<f$)
\begin{equation}
    \label{eq:integral}
    \Gamma_{f,j} \approx \frac{\Omega^2}{\Delta_j} \left(\frac{J}{\Delta_j} \right)^{|j-f|} .
\end{equation}
This shows that the cavity-mediated hopping decays exponentially with the distance $|j-f|$, with characteristic length scale $\xi_{j} = \log (|\Delta_j|/J)$ corresponding to the localization length (see Fig.~\ref{fig:coupling}). The sign of the coupling is determined by the sign of $\Delta_j$; in particular, for $\Delta_j < 0$ the coupling alternates in sign according to $\Gamma_{f,j} \sim (-1)^{1+j-f}$. Within the perturbative regime, the eigenstates of the lower band therefore remain localized. Although the associated localization length is independent of the light--matter coupling strength in this regime, it increases with the hopping amplitude $J$ in the excited band. In the limiting case $J \to 0$, one recovers $\xi_{j}\to 0$, as expected for trivially localized orbitals. 

To investigate the behavior beyond the perturbative regime, whose validity is restricted to $\Omega/|\Delta_j| \ll 1$, we next employ exact diagonalization of the full Hamiltonian to gain insight into the structure of the eigenmodes. We then complement this analysis with a transfer-matrix approach, which provides an efficient way to determining the localization properties in the thermodynamic limit.

\begin{figure*}[htbp!] 
\centering
\includegraphics[width=\textwidth]{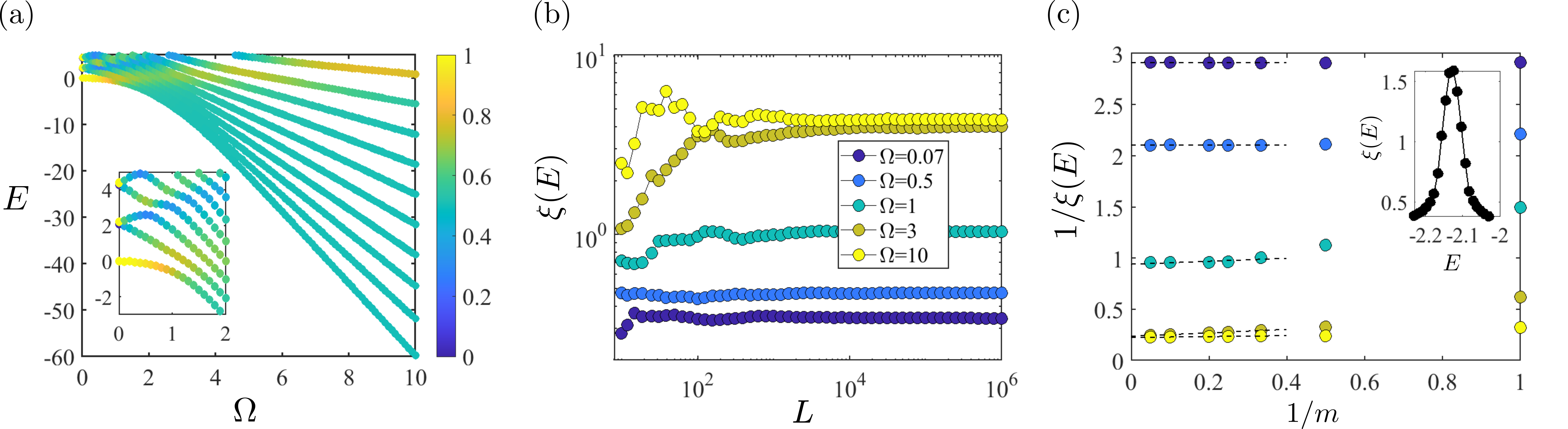}
\caption{\textbf{Enhancement of the localization length in the two-band model.} (a) Energy dispersion of the two-band model coupled to the cavity in the single-electron sector as a function of the coupling strength $\Omega$ for system length $L=10$ and keeping $m=20$ bosonic Fock states; the inset shows a zoomed-in region. The colorbar indicates the weight of the eigenstates on the lower band (with any 
photon number). (b) Localization length $\xi(E)$, obtained via a transfer-matrix calculation, as a function of the system size $L$ for several coupling strengths $\Omega$ and fixed $m=20$. For each value of $\Omega=0.007,0.5,1,3,10$, the energy is chosen at the center of the lowest band, namely $E=-2.3\times10^{-5},-0.12,-0.58,-7.68,-59.76$, respectively.
(c) Inverse localization length plotted against $1/m$ for different values of $\Omega$ [same color code and energies as in panel (b)]; Dashed lines represent linear fits to the data for $m=3,4,5,10,20$. Inset: $\xi$ as a function of the energy $E$ for $\Omega=3$. A small nearest-neighbor hopping $J_g=0.001$ is included in the lower band to facilitate the numerical analysis. Parameters are $W=0.1$, $J=0.02$, $\epsilon_e=1$, and $\omega=1.1$. The lower-band onsite energies $\epsilon_{g,i}$ are drawn from the interval $[-W/2,W/2]$, which corresponds to detunings $\Delta_{i} \in [-2.15, -2.05]$.}
    \label{fig:tmatrix-energy}
\end{figure*}




\subsection{Exact diagonalization}
\label{exact_diag}

As discussed above and in Fig.~\ref{fig:sketch1}, the structure of the states in our two-band model is as follows. The state $\ket{g,j}\otimes\ket{0}$ (disordered band with zero photons) is coupled, via the counter-rotating terms $c^\dagger_{e,j} c_{g,j} a^\dagger$ in the coupling Hamiltonian $V$, to the state $\ket{e,j}\otimes\ket{1}$ (excited band with one photon). The latter is coupled both to its nearest neighbors $\ket{e,j\pm1}\otimes\ket{1}$ within the same band and to the state $\ket{g,j}\otimes\ket{2}$ (disordered band with two photons) through the co-rotating terms $c^\dagger_{g,j} c_{e,j} a^\dagger$. This generates a ladder of hybrid states with increasing photon number, clearly visible in the exact-diagonalization spectrum of $H$ shown in Fig.~\ref{fig:tmatrix-energy}(a). For small $\Omega$, the spectrum consists of a disordered zero-photon band at energy $0$, an extended one-photon band at $\epsilon_e+\omega$, a two-photon band at $2\omega$, and so on, which corresponds to the perturbative regime. As $\Omega$ increases and eventually becomes comparable to the spectral gap $\Delta_{j}$, a regime that may be regarded as ultrastrong coupling at the single-electron level, these bands hybridize and give rise to eigenstates carrying approximately equal weight in the lower and excited electronic states for any photon number. The scaling of the ground-state energy and band gaps with $\Omega$ can be understood by mapping the model onto the Dicke Hamiltonian~\cite{EmaryBrandes2003}, obtained by introducing the two-level raising operators $\sigma_j^{+} = c^\dagger_{e,j} c_{g,j}$. 

Having characterized the spectral structure of the hybrid light--matter eigenstates in the two-band model, we now turn to the investigation of their localization properties beyond the perturbative regime.

\subsection{Transfer Matrix Approach}
\label{transfer_mat}

In this section we aim to assess whether the system remains localized or not and if yes compute the localization length for any coupling strength $\Omega$. For that purpose we use a transfer matrix approach, which has been successfully used to determine Anderson localization in different lattice configurations~\cite{pichard_finite_1981,chalker1993scattering}. Truncating the photon number to $m-1$, the state space forms an $m \times L$ lattice. Thus, the wavefunction of a single particle with energy $E$ at the position $(j,\alpha)$ on this 2D lattice, with $(j,\alpha) \in [1,L] \times [1,m]$ satisfies the Schrödinger equation
\begin{equation}
    E \psi^{(j)}_{\alpha} = \sum_{\beta=1}^m V^{(j)}_{\alpha\beta} \psi^{(j)}_{\beta} + \sum_{\beta=1}^m t_{\alpha\beta} \left( \psi^{(j+1)}_{\beta} + \psi^{(j-1)}_{\beta} \right)~,
\end{equation}
where $V^{(j)}$ is an $m \times m$ Hermitian matrix representing the intra-site transitions between lower and upper bands and different photon numbers at site $j$, and $t$ is an $m \times m$ diagonal matrix representing the inter-site nearest-neighbor hopping terms. The wavefunction can be constructed using the $2m \times 2m$ transfer matrix
\begin{equation}
T_j (E) = \left( \begin{array}{c|c}
        t^{-1} \left(E~\mathbf{I} - V^{(j)}\right) & - \mathbf{I} \\ \hline
        \mathbf{I} & 0
    \end{array} \right),   
\end{equation}
where $\mathbf{I}$ is the $m \times m$ identity matrix, as 
\begin{align}
    \left( \begin{array}{c}
        \psi^{(j+1)}_{1} \\ 
        \vdots \\
        \psi^{(j+1)}_{m} \\ 
        \psi^{(j)}_{1} \\
        \vdots \\
        \psi^{(j)}_{m}
    \end{array} \right) = T_j(E) \left( \begin{array}{c}
        \psi^{(j)}_{1} \\ 
        \vdots \\
        \psi^{(j)}_{m} \\ 
        \psi^{(j-1)}_{1} \\
        \vdots \\
        \psi^{(j-1)}_{m}
    \end{array} \right) ~.
\end{align}
Since some components of $V^{(j)}$ are random, the eigenstates with energy $E$ are reconstructed from left to right as the product of random matrices $T_{L-1}(E)  \cdots T_3 (E)T_2(E)$,
\begin{equation}
    \left( \begin{array}{c}
        \psi^{(L)}_{1} \\ 
        \vdots \\
        \psi^{(L)}_{m} \\ 
        \psi^{(L-1)}_{1} \\
        \vdots \\
        \psi^{(L-1)}_{m}
    \end{array} \right) = T_{L-1} \dots T_3 T_2  \left( \begin{array}{c}
        \psi^{(2)}_{1} \\ 
        \vdots \\
        \psi^{(2)}_{m} \\ 
        \psi^{(1)}_{1} \\
        \vdots \\
        \psi^{(1)}_{m}
    \end{array} \right)   .
\end{equation}
Thus, to evaluate the decay of the amplitude of $\psi_\alpha^{(L)}$ with $L$, we need to look at the eigenvalues of the matrix product $T_{L-1}  \dots T_3 T_2$. Importantly, $T_j$ is conjugated to a symplectic matrix
\begin{equation}
   \tilde{T}_j = \left( \begin{array}{c|c}
       t^{1/2} & 0 \\ \hline
       0 & t^{-1/2}
   \end{array} \right) T_j \left( \begin{array}{c|c}
       t^{1/2} & 0 \\ \hline
       0 & t^{-1/2}
  \end{array} \right)^{-1} ~,
\end{equation}
so the eigenvalues of $T_{L-1}  \dots T_3 T_2$ are the same as the ones of the symplectic matrix product $\tilde{T}_{j}  \dots \tilde{T}_3 \tilde{T}_2$. The spectrum of any symplectic matrix is of the form $$\left\{ \lambda_1 , \lambda_2, \dots, \lambda_m, \lambda^{-1}_1, \lambda^{-1}_2, \dots,\lambda^{-1}_m \right\},$$ with eigenvalues that occur in reciprocal pairs, and $$|\lambda_1| \geq |\lambda_2| \geq \dots \geq |\lambda_{m}|.$$ Thus, to evaluate the localization length we can restrict to the part of the spectrum with $|\lambda_\alpha| \geq 1$. Oseledets' multiplicative ergodic theorem~\cite{oseledec1968multiplicative} ensures that the eigenvalues of the matrix product converge when $L \rightarrow \infty$, which can be interpreted as a self-averaging property. The corresponding Lyapunov exponents are $\log | \lambda_{\alpha} |$, for $ \alpha= 1, \dots , m$, and the different localization lengths for an energy $E$ are then $\xi_{\alpha}(E) = ( \log | \lambda_{\alpha} | )^{-1}$. We thus define the localization length of the system at energy $E$ as $\xi(E) = \max_{\alpha \in [1,m]} \{ \xi_{\alpha}(E) \}$.


In Fig.~\ref{fig:tmatrix-energy}(b), we display the localization length $\xi(E)$ as a function of the system size $L$ for fixed energies $E$ located at the center of the lowest-energy band, and for several light--matter coupling strengths in the narrow-band regime $2J \ll |\Delta_i|$. For weak couplings $\Omega < \Delta_i$, the localization length rapidly converges to a value smaller than one lattice site, consistent with the perturbative analysis. For stronger couplings $\Omega > |\Delta_i|$, convergence requires system sizes as large as $L \gtrsim 10^4$, yet the localization length remains finite in the thermodynamic limit. Importantly, although finite, we find that the localization length is strongly enhanced and extends over several lattice sites in the single-electron ultrastrong-coupling regime $\Omega \gg |\Delta_i|$. Convergence with respect to the photon number $m$, shown in Fig.~\ref{fig:tmatrix-energy}(c), likewise yields finite limiting values of $1/\xi(E)$, which themselves stabilize as $\Omega$ increases. Furthermore, the localization length exhibits a smooth dependence on the energy $E$, as shown in the inset of Fig.~\ref{fig:tmatrix-energy}(c), and it converges to an $L$-independent curve in the limit of large system sizes. 

In conclusion, we find that the states of the disordered band remain localized in the thermodynamic limit. However, the localization length increases with the light--matter coupling strength and eventually saturates at a value of a few lattice sites in the single-electron ultrastrong-coupling regime. In the following section, we investigate the localization properties of the model introduced in Ref.~\cite{ciuti_cavity-mediated_2021}, which constitutes a natural analogue of the two-band minimal model considered previously, with particular emphasis on the spatial decay of the effective hopping amplitude.

\begin{figure*}
(a) \hspace{5cm} (b) \hspace{5cm} (c)
\centering
\includegraphics[width=\textwidth]{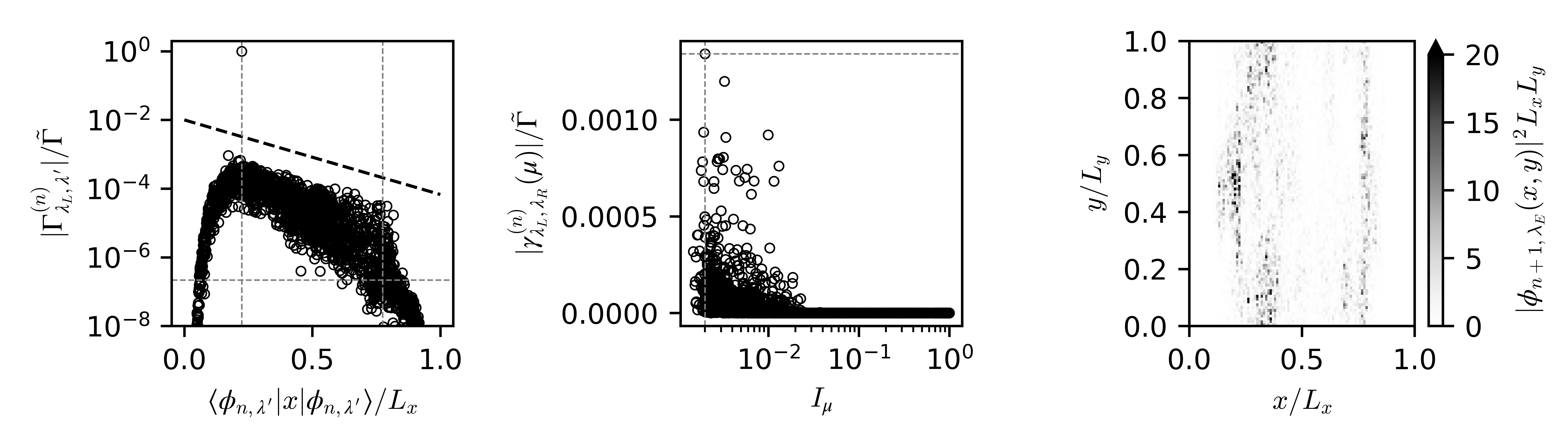}
\caption{Exponentially-decaying hopping in a disordered Landau band. (a) Normalized effective coupling Eq.~\eqref{effective_coupling} between an eigenstate $\ket{\phi_{n,\lambda_L}}$ on the left edge and all the other eigenstates $\ket{\phi_{n,\lambda'}}$ for the Landau band Hamiltonian, plotted as a function of the average position $\bra{\phi_{n,\lambda'}} x \ket{\phi_{n,\lambda'}}$ of each eigenstate. Thick dashed line: exponential decay of the coupling (guide for the eye). Thin dashed lines: average positions of $\ket{\phi_{n,\lambda_L}}$ on the left edge, a representative state $\ket{\phi_{n,\lambda_R}}$ on the right edge, and the corresponding normalized coupling. (b) Contributions of each eigenstate $\ket{\phi_{n+1,\mu}}$ to the normalized effective coupling between $\ket{\phi_{n,\lambda_L}}$ (on the left edge) and $\ket{\phi_{n,\lambda_R}}$ (on the right edge), shown versus their inverse participation ratio. Thin dashed lines: state $\ket{\phi_{n+1,\lambda_E}}$  with the maximal contribution.
(c) Normalized real-space probability density function of the state $\ket{\phi_{n+1,\lambda_E}}$ that gives the maximal contribution to the coupling $\Gamma_{\lambda_L,\lambda_R}^{(n)}$ between eigenstates on the opposite edges. Same parameters as in Ref.~\cite{ciuti_cavity-mediated_2021}: $n=4$, $N=2400$, $L_{x}=10\mu$m, $L_{e}=2.5\mu$m, $r=0.9$, $W_{0}=0.03 \omega_{0}$, $N_{\textrm{imp}}=2000$, $\mathcal{V}^{(\textrm{imp})}_{j} \in [-1.5, 1.5]\times 10^{-5}\omega_{0}L_{x}L_{y}$, $\omega_{0}=2.09$THz, $g_{0}=0.0031\omega_{0}$.}
\label{fig:LLs}
\end{figure*}

\section{Disordered Landau band}
\label{Landau_band}

We now turn to the model originally introduced in Ref.~\cite{ciuti_cavity-mediated_2021}, describing a two-dimensional gas of spinless electrons on an annulus of circumference $L_y$ and of width $L_x$ subjected to a static magnetic field. We start by briefly reviewing the model and the main results of Ref.~\cite{ciuti_cavity-mediated_2021}.

In the Landau gauge, translational invariance is preserved along the $y$ direction, so the single-particle eigenstates are characterized by two quantum numbers: the Landau-level index $n = 0,1,2,\ldots$ and the $y$-momentum $k=2\pi p /L_{y}$, which determines the guiding--center coordinate in the $x$ direction, $k l^2_{0}$. Here $p = 1,2,\ldots, N$, with $N = L_x L_y / (2\pi l_0^{2})$ the Landau-level degeneracy and $l_0$ the magnetic length. The corresponding wavefunctions are $\varphi_{n,k}(\mathbf{r}) = \langle\mathbf{r}|\varphi_{n,k}\rangle = \chi_n(x - k l_0^{2})\, e^{i k y}/\sqrt{L_y}$, where $\chi_n(x)$ denotes the normalized harmonic-oscillator wavefunctions. To model confinement near the sample edges, we introduce a wall potential of width $L_e \ll l_0$ defined as $W(x)=W_0 \tan^{2}[r\pi(x-L_e)/(2L_e)]$ for $0 < x < L_e$, and $W(x)=W_0 \tan^{2}[r\pi(x-L_x+L_e)/(2L_e)]$ for $L_x - L_e < x < L_x$, while $W(x)=0$ in the bulk, with $r \simeq 1$. Disorder is incorporated through $N_{\mathrm{imp}}$ point-like impurity potentials $V(\mathbf{r})=\sum_{j=1}^{N_{\mathrm{imp}}} \mathcal{V}^{(\mathrm{imp})}_{j}\,\delta(\mathbf{r}-\mathbf{r}_{j})$, where the amplitudes $\mathcal{V}^{(\mathrm{imp})}_{j}$ and positions $\mathbf{r}_j$ are independently drawn from uniform distributions. These impurities break translational invariance and thus induce scattering between different guiding-center states $k$. The disorder strength is taken to be small compared to the cyclotron energy $\omega_{0}$ so that Landau-level mixing remains negligible. The electronic Hamiltonian reads 
\begin{equation}
\label{el_H_LL}
H_{\textrm{el}} = \sum_{n} \sum_{k,k'} \left[(E_{n} + W_{k})\delta_{k,k'} + V_{k,k'} \right]c^{\dagger}_{n,k} c^{\phantom{\dagger}}_{n,k'}~, 
\end{equation}
where $E_{n}=n\omega_{0}$, $W_{k}=\int \! d\mathbf{r} \, \varphi^{*}_{n, k} (\mathbf{r}) W(x) \varphi_{n, k} (\mathbf{r}) \simeq W (kl^2_{0})$ and $V_{k,k'}=\sum_{j=1}^{N_{\textrm{imp}}} \mathcal{V}^{(\textrm{imp})}_{j} \varphi^{*}_{n, k} (\mathbf{r}_{j}) \varphi_{n, k'} (\mathbf{r}_{j})$. The electronic Hamiltonian~\eqref{el_H_LL} can be diagonalized as $H_{\textrm{el}} = \sum_{n,\lambda} \epsilon_{n,\lambda}\, d^{\dagger}_{n,\lambda} d_{n,\lambda}$, where $\epsilon_{n,\lambda}$ are the energies of the Landau-band orbitals and $\ket{\phi_{n,\lambda}}$ the corresponding eigenvectors.

Then one introduces the dipolar coupling between two consecutive Landau bands and a single-mode cavity field. Denoting by $a^{\dagger}$ ($a$) the creation (annihilation) operator of a cavity photon dressed by its interaction with the electron density and having energy $\omega$, the total Hamiltonian is written as~\cite{ciuti_cavity-mediated_2021} 
\begin{align*}
H &= \sum_{n,\lambda} \epsilon_{n,\lambda} d^\dagger_{n,\lambda} d_{n,\lambda} + \omega a^\dagger a^{\phantom{\dagger}} \\
&+ \sum_{n}\sum_{\lambda,\mu} g^{(n,n+1)}_{\lambda,\mu} ( a^{\phantom{\dagger}}+a^{\dagger}) d^\dagger_{n+1,\mu} d^{\phantom{\dagger}}_{n,\lambda} + {\rm h.c.}~,   
\end{align*}
where $g_{\lambda,\mu}^{(n,n')} = g_{0}\sqrt{n'}\sum_{k} \langle \phi_{n',\mu} | \varphi_{n',k} \rangle \langle \varphi_{n,k} | \phi_{n,\lambda} \rangle$ is the coupling between the $\lambda$th eigenstate of band $n$ and the $\mu$th eigenstate of band $n'$, and $g_{0}$ is a coupling constant determined by microscopic parameters. 
Ref.~\cite{ciuti_cavity-mediated_2021} focuses on the cavity-mediated hopping between two eigenstates $\lambda$ and $\lambda'$ within the same Landau band $n$, which using second-order perturbation theory reads
\begin{subequations}
\label{effective_coupling}
\begin{equation}
    \Gamma_{\lambda,\lambda'}^{(n)} = \sum_{\mu} \gamma_{\lambda,\lambda'}^{(n)}(\mu)
\end{equation}
where
\begin{equation}
 \gamma_{\lambda,\lambda'}^{(n)}(\mu) = \frac{g_{\lambda,\mu}^{(n,n+1)}\, g_{\mu,\lambda'}^{(n+1,n)}}{\epsilon_{n,\lambda} - \epsilon_{n+1,\mu} - \omega}~.
\end{equation}
\end{subequations}
Ref.~\cite{ciuti_cavity-mediated_2021} showed that the cavity-mediated hopping (\ref{effective_coupling}) is non-zero.

In Fig.~\ref{fig:LLs}(a) we reproduce these results: we show the cavity-mediated hopping between an eigenstate localized on the left edge and all the other eigenstates as a function of the average position of the states. The effective coupling is normalized by 
\begin{equation}
\tilde{\Gamma}=\frac{g^2_{0}\sqrt{n+1}}{\omega_{0}+\omega}.
\end{equation}
We stress that the effective coupling decays exponentially with the distance from the edge, consistent with the localized nature of the states in the lowest disordered Landau band. However, the localization length in the $x$ direction can be quite large. The localization length extracted from Fig.~\ref{fig:LLs}(a) is approximately $2\,\mu\mathrm{m}$, much larger than the magnetic length $l_0 \approx 30\,\mathrm{nm}$ (see caption for the parameters used) and therefore than the characteristic localization length of edge states in the tails of the Landau band~\cite{halperin_quantized_1982}. This behavior closely parallels the results obtained in the two-band minimal model. We attribute the strong enhancement of the localization length to the fact that the effective coupling between localized states in the lower Landau band is mediated by extended states in the upper Landau band. This is shown in Fig.~\ref{fig:LLs}(b), where we analyze the contributions $\gamma_{\lambda_L,\lambda_R}^{(n)}(\mu)$ of each eigenstate $\mu$ to the coupling between two states localized on the left ($\lambda_L$) and right edge ($\lambda_R$). The contributions are plotted against the inverse participation ratio of the corresponding eigenstate
\begin{equation}
    \label{IPR}
    I^{(n)}_\mu = \sum_k \vert \langle \varphi_{n+1,k}\vert \phi_{n+1,\mu} \rangle |^4~,
\end{equation}
which quantifies localization in the eigenbasis in the absence of disorder.
More specifically, $I_\mu\sim\mathcal{O}(1)$ indicates a localized state, while $I_\mu\sim\mathcal{O}(1/N)$ denotes an extended state. Figure~\ref{fig:LLs}(b) shows that the most extended states provide the dominant contributions to the coupling between the two edges. In Fig.~\ref{fig:LLs}(c), we also plot the real-space probability density associated to the wavefunction $\ket{\phi_{n+1,\lambda_E}}$ in the excited Landau band and for the eigenstate $\lambda_E$ that gives the largest contribution to the inter-edge coupling. The spatially extended character of this probability density, associated with a localization length comparable to the system size $L_x$, demonstrates that the corresponding eigenstate $\lambda_E$ is extended not only along the $y$ direction, where translational invariance is preserved, but also along the $x$ direction.

\section{Conclusion}
\label{conc}

In conclusion, we have studied a two-band model with localized states in the lowest energy band and delocalized excited states coupled to a single cavity mode in the regime where the hopping in the excited band remains much smaller than the lowest energy gap of the spectrum. Using a perturbative analysis valid when this energy gap is also much larger than the light--matter coupling, we showed that the cavity-assisted coupling between two states of the disordered band located at different sites remains short-ranged, but scales quadratically with the light--matter coupling strength $\Omega$ and exponentially with a localization length that itself increases with the hopping amplitude in the excited band. A transfer-matrix analysis further shows that all eigenstates remain localized even at stronger light--matter coupling strengths. Interestingly, outside the perturbative regime, the localization length becomes strongly enhanced with increasing coupling strength and eventually saturates at a value of a few lattice sites in the single-electron ultrastrong-coupling regime, where the on-site light--matter coupling exceeds the lowest energy gap. 

We argued that this two-orbital model in the small bandwidth regime is relevant for disordered quantum Hall systems that were studied in previous literature. In this model, we have found that the effective coupling between two localized states in the lowest Landau band is mostly mediated by states in the excited Landau band that are extended in the bulk of the system. Although the effective coupling between edge states in the lowest Landau band reported in Ref.~\cite{ciuti_cavity-mediated_2021} still decays exponentially with distance, it is characterized by a localization length that is much larger than the microscopic length scale of a bare Landau band in the absence of cavity dressing and can reach several microns. We emphasize that this regime is particularly relevant for micron-scale mesoscopic systems, in which such states may effectively appear delocalized.

Future directions include extending our two-band model to a fully two-dimensional lattice, yielding an effective three-dimensional problem once cavity photons are included, where the fate of eigenstate localization remains an open question, reminiscent of the original Anderson model~\cite{Anderson1958,Abrahams1979}. Furthermore, extending the transfer-matrix approach to compute generalized inverse participation ratios 
could offer deeper insight into the system's localization properties, particularly in view of recent reports of multifractality in disordered light--matter--coupled systems~\cite{PhysRevA.105.023714,PhysRevB.109.064202,PhysRevA.109.033716,gt7l-pyql}.

\begin{acknowledgments}
We are indebted to C.~Ciuti for taking the time to perform calculations and share results with us, and for pointing out an important mistake in the first version of this manuscript. We also thank D. Bori\c{c}i, A. Vasanelli and C. Sirtori for useful discussions. This work has received funding from the French National Research Agency under the Investments of the Future Program projects ANR-21-ESRE-0032 (aQCess), ANR-22-CE47-0013-02 (CLIMAQS), ANR-23-CE30-0022-02 (SIX), and ANR23-PETQ-0002 (QUTISYM) and from the European Union via the program HORIZON-MSCA-2023-DN-01 101169225 (SPARKLE). 
\end{acknowledgments}

\bibliographystyle{jabbrv_apsrev4-1}
\bibliography{refs.bib}

\end{document}